\newif\iflandscape
\newif\ifportrait
\portraittrue
\newlength{\extralineskip}
\ifportrait
  \documentstyle[12pt]{article}
\fi
\iflandscape
  \documentstyle[twocolumn]{article}
\fi
\def\tr#1{{\rm tr}\kern-3pt\left[#1\right]}

\def\bea{\begin{eqnarray}}
\def\eea{\end{eqnarray}}
\def\nn{\nonumber}

\def\beq{\begin{equation}}
\def\eeq{\end{equation}}
\def\ba{\beq\new\begin{array}{c}}
\def\ea{\end{array}\eeq}
\def\be{\ba}
\def\ee{\ea}

\def\f{1\over}
\makeatletter
\newdimen\normalarrayskip              % skip between lines
\newdimen\minarrayskip                 % minimal skip between lines
\normalarrayskip\baselineskip
\minarrayskip\jot
\newif\ifold             \oldtrue            \def\new{\oldfalse}
\def\arraymode{\ifold\relax\else\displaystyle\fi} % mode of array enrties
\def\eqnumphantom{\phantom{(\theequation)}}     % right phantom in eqnarray
\def\@arrayskip{\ifold\baselineskip\z@\lineskip\z@
     \else
     \baselineskip\minarrayskip\lineskip2\minarrayskip\fi}
\def\@arrayclassz{\ifcase \@lastchclass \@acolampacol \or
\@ampacol \or \or \or \@addamp \or
   \@acolampacol \or \@firstampfalse \@acol \fi
\edef\@preamble{\@preamble
  \ifcase \@chnum
     \hfil$\relax\arraymode\@sharp$\hfil
     \or $\relax\arraymode\@sharp$\hfil
     \or \hfil$\relax\arraymode\@sharp$\fi}}
\def\@array[#1]#2{\setbox\@arstrutbox=\hbox{\vrule
     height\arraystretch \ht\strutbox
     depth\arraystretch \dp\strutbox
     width\z@}\@mkpream{#2}\edef\@preamble{\halign \noexpand\@halignto
\bgroup \tabskip\z@ \@arstrut \@preamble \tabskip\z@ \cr}%
\let\@startpbox\@@startpbox \let\@endpbox\@@endpbox
  \if #1t\vtop \else \if#1b\vbox \else \vcenter \fi\fi
  \bgroup \let\par\relax
  \let\@sharp##\let\protect\relax
  \@arrayskip\@preamble}
\def\eqnarray{\stepcounter{equation}%
              \let\@currentlabel=\theequation
              \global\@eqnswtrue
              \global\@eqcnt\z@
              \tabskip\@centering
              \let\\=\@eqncr
              $$%
 \halign to \displaywidth\bgroup
    \eqnumphantom\@eqnsel\hskip\@centering
    $\displaystyle \tabskip\z@ {##}$%
    &\global\@eqcnt\@ne \hskip 2\arraycolsep
         $\displaystyle\arraymode{##}$\hfil
    &\global\@eqcnt\tw@ \hskip 2\arraycolsep
         $\displaystyle\tabskip\z@{##}$\hfil
         \tabskip\@centering
    &{##}\tabskip\z@\cr}
\makeatother
\def\f{1\over }

\begin{document}
\begin{titlepage}
\setcounter{footnote}0
\begin{center}
%{\it P.N.Lebedev Institute preprint}
\hfill ITEP M-8/93\\
\hfill FIAN/TD-22/93\\
\hfill CRM-1934\\
\hfill hep-th/9312213\\
\vspace{0.3in}
{\LARGE\bf On a $c$-number quantum $\tau$-function}
\footnote{Contribution to the Proceedings of III International
Conference on Mathematical Physics, String Theory and Quantum Gravity,
Alushta, June, 1993}
\\[.4in]
{\Large A. Mironov\footnote{E-mail address:
mironov@td.fian.free.net, mironov@grotte.teorfys.uu.se}}\\
\bigskip {\it Theory Department,  P. N. Lebedev Physics
Institute, Leninsky prospect, 53,\\ Moscow,~117924, Russia}
\bigskip
\\
{\Large A. Morozov\footnote{E-mail address: morozov@vxitep.itep.msk.su,
morozov@vxdesy.desy.de}}\\
\bigskip {\it ITEP, Moscow, 117 259, Russia}
\bigskip
\\
{\Large Luc Vinet\footnote{E-mail address: vinet@ere.umontreal.ca,
vinet@lps.umontreal.ca}}\\
\bigskip {\it Centre de recherches math\'ematiques,
         Universit\'e de Montr\'eal,\\
         C. P. 6128-A, Montr\'eal (Qu\'ebec) H3C 3J7}
\end{center}
\bigskip
\bigskip

\centerline{\bf ABSTRACT}
\begin{quotation}
We first review the properties of the conventional $\tau$-functions of the KP
and Toda-lattice hierarchies. A straightforward generalization is then
discussed. It corresponds to passing from differential to finite-difference
equations; it does not involve however the concept of operator-valued
$\tau$-function nor the one associated with non-Cartanian (level $k\ne1$)
algebras. The present study could be useful to understand better $q$-free
fields and their relation to ordinary free fields.
\end{quotation}
\end{titlepage}
\clearpage

\newpage

\section{Introduction}
\setcounter{footnote}{0}
The $\tau$-functions originally associated
with integrable Toda-lattice hierarchies, have been identified as sections of
determinant bundles over the universal
moduli space \cite{Jap,SW}, and also seen to admit a group-theoretical
interpretation \cite{Kac}. They are now attracting attention again
because of their appearance as
non-perturbative partition functions in quantum field/string
theory \cite{BDGKMS}.
This is giving a new stimulus to the search of a broader
interpretation of these $\tau$-function which would go
beyond free-fermion models and commuting differential
flows \cite{UFN}.  The two most obvious directions of
generalization consist in moving from free fermions to
generic conformal models on one hand and from continuous 2-dimensional
spectral surfaces to their quantum and/or discrete counterparts on the other
hand.
In the group-theoretical approach this corresponds to going
beyond the level $k=1$ Kac-Moody algebras and from ``classical'' to
quantum groups respectively.

The purpose of this note is to discuss briefly some preliminary
material, concerning the ``quantum $\tau$-functions'' that follow from the
second direction of generalization. We concentrate here on a very
particular aspect of the corresponding theory, i.e.\ on a
possible relation of these ``quantum $\tau$-function'' to
the conventional ``classical'' Toda-lattice $\tau$-function
interpreting the former one as a $c$-numbers. We proceed by rephrasing results
obtained in refs.~\cite{Sat}, where solutions to finite-difference analogues
of Hirota equations are given. These solutions are found for the ``first''
time-variable. We present here one particular way to
introduce the other ``times''. In this construction the $c$-number ``quantum
$\tau$-function'' is obtained from the ``classical'' one, merely through a
change of variable
which appears to be a kind (albeit special) of Miwa transformation. Within
this framework there is no need to
go beyond the standard Segal-Wilson infinite Grassmannian and the standard
free fermions (no $q$-fermions are required). Relations to
generic operator-valued objects as well as to other aspects of the theory
of quantum $\tau$-functions
\cite{q-tau} will be discussed elsewhere.

\section{Conventional $\tau$-function}

We begin by reminding the reader of the definition of the conventional
$\tau$-function for the Toda-lattice and KP hierarchies. There are
actualy several more or less equivalent definitions. Any
generalization should prove natural irrespective of which definition is
adopted. For the sake of completeness
and future use we give below the whole list. In this note however, we will
deal mostly with only one of these definitions.

The conventional $\tau$-function can be characterized by any of
the following properties:

1) It satisfies a bilinear Hirota equation, which for the KP hierarchy
looks like
\be
\oint \frac{dz}{z} V^+\{z|t\}\otimes V^-\{z|t'\}   \tau\{t\}\otimes \tau\{t'\}
= 0,
\label{Hirota} \\
V^\pm\{z|t\} = \exp
\left(\pm\sum_{n>0}\frac{1}{nz^n}\frac{\partial}{\partial t_n}\right).
\label{V-op}
\ee

2) It can be written as a correlator of free fermions,
\be
\tau_N\{t,\bar t\} = \langle N | e^{H\{t\}} {\cal G} e^{\bar H\{\bar t\}}
|N\rangle,
\label{fftau}
\ee
where
\be
H\{t\} = \sum_{n>0} t_nJ_{+n}, \ \ \ \bar H\{\bar t\} =
\sum_{n>0} \bar t_n J_{-n}, \nn \\
J(z) = \sum_{n=-\infty}^{\infty} J_nz^{n-1} = \tilde
\psi(z)\psi(z) = \partial\phi(z), \ \ \
{\cal G} = \exp \left(\sum_{m,n}{\cal
A}_{mn}\tilde\psi_n\psi_n\right), \nn \\
\psi(z)=\sum_{i\in Z}\psi_i z^i,\ \ \tilde \psi(z)=\sum_{i\in Z}
\tilde\psi_iz^{-i},\\
\tilde \psi_{n}|N\rangle = 0,
\ \ \hbox{for}\ \ n\ge N, \ \psi_{n}|N\rangle = 0 \ \ {\rm for}\ \ n<N\ .
\label{ffdefs}
\ee

3) In Miwa coordinates, the KP $\tau$-function  can be represented in the
determinant form:
\be
\tau\{t\} = \frac{\det \Psi_a(\lambda_b)}{\Delta (\lambda)},
\label{Miwadet}
\ee
where
\be
t_n = t^{(0)}_n + \frac{1}{n}\sum_{\alpha} \lambda_\alpha^{-n},
\bar t_n = \bar t^{(0)}_n + \frac{1}{n}\sum_{\alpha}
\bar \lambda_\alpha^{n},
\label{Miwatimes}
\ee
and $\Delta(\lambda) = \prod_{\alpha>\beta}(\lambda_\alpha -
\lambda_\beta)$.
The analogous representation of the generic Toda-lattice
$\tau$-function is a little more sophisticated, see
refs.~\cite{UFN,Rev,char} for details. Sometimes it is convenient to
interpret the $\lambda_a$'s as eigenvalues of some matrix $\Lambda$
\cite{UFN,Rev}.

The Miwa coordinates can be used to introduce a sort of Fourier
transform with respect to the variables $\{t,\bar t\}$,
\be
\tau_N\{t,\bar t\} = \hat\tau\{p(\lambda),\bar p(\bar\lambda)\},
\label{p-tau}
\ee
with
\be
t_n = t^{(0)}_n + {\f n}\int_{d\lambda} p(\lambda) \lambda^{-n}, \ \ \
\bar t_n = \bar t^{(0)}_n + {\f n}\int_{d\bar\lambda}
\bar p(\bar\lambda) \bar\lambda^{n}.
\label{t-p}
\ee
If $p(\lambda)=\sum_ap_a\delta(\lambda-\lambda_a)$ and all non-vanishing
$p_a=1$, we recover the parameterization (\ref{Miwatimes}).

In terms of $\hat\tau(p_a,\bar p_a)$ the Hirota equations become
finite-difference equations \cite{hir-dis}:
\be
 (\lambda_a-\lambda_b)\hat\tau_N(p_a+1,p_b+1;p_c)
\hat\tau_N(p_a,p_b;p_c+1)+\\+(\ \hbox{cyclic permutations of}\ a,\ b,\ c)=0\
\ \hbox{for any three}\ \lambda_a,\ \lambda_b,\ \lambda_c;\label{hirdis1}
\ee
\be
\hat\tau_N(p+1,\bar p)\hat\tau_N(p,\bar p+1)-\hat\tau_N(p+1,\bar p+1)
\hat\tau_N(p,\bar p)=\\={\bar\lambda_b \over \lambda_a}
\hat\tau_{N+1}(p+1,\bar p+1)
\hat\tau_{N-1}(p,\bar p)\ \
  \hbox{for any two}\ \lambda_a,\ \lambda_b;\\
\ldots
\label{hirdis2}
\ee
These equations are an immediate consequence of (\ref{V-op}). Actually,
there is an infinite set of equations, but it is sufficient to consider
only (\ref{hirdis1}) in order to fix the KP flows associated with positive
times, the conjugate equation is
necessary to fix the KP flows w.r.t.\ negative
times. Moreover, it is enough to
consider only (\ref{hirdis2})
in order to obtain the complete Toda-lattice hierarchy.

4) Most relevant to our discussion is
another determinant formula for the Toda-lattice $\tau$-function
\cite{UFN,Rev}:
\be
\tau_N\{t,\bar t\} = \det_{-\infty < i,j \le N} H_{ij}\{t,\bar t\},
\label{infdet}
\ee
where the matrix $H_{ij}$ is characterized by the property,
\be
\frac{\partial}{\partial t_k} H_{ij}= H_{i+k,j} =
\left( \frac{\partial}{\partial t_1}\right)^k H_{ij}; \\
\frac{\partial}{\partial \bar t_k} H_{ij}= H_{i,j+k} =
\left( \frac{\partial}{\partial \bar t_1}\right)^k H_{ij}. \\
\label{difshift}
\ee
These equations can be solved by a sort of Fourier transform, and
their general solution is provided by an arbitrary measure function
$\mu(z,\bar z)$:
\beq
H = \int_{dzd\bar z} \left(\prod_{k=1}^\infty e^{t_k z^k}\right)
\mu(z,\bar z)
\left(\prod_{k=0}^\infty e^{\bar t_k \bar z^k}\right),\ \
T_{lm} = \int_{dzd\bar z} z^l\mu(z,\bar z)\bar z^m,
 \eeq
which implies that
\be
H_{ij}\{t,\bar t\} = \sum_{l,m} P_{l-i}\{t\} T_{lm}
P_{m-j}\{\bar t\}.
\label{H-shur}
\ee
$T_{lm}$ is already independent of $t$ and $\bar t$ (in fact, $T_{lm}$
describes the adjoint action of the element of the
Grassmannian ${\cal G}$ on fermionic modes). The
Schur polynomials $P_m\{t\}$ are defined by
\be
\prod_{k=0}^{\infty} e^{t_kz^k} = \sum_{m=0}^{\infty}P_m\{t\} z^m,
\label{shurdef}
\ee
so that
\be
\frac{\partial}{\partial t_k}P_m\{t\} = P_{m-k}\{t\} = \left(
\frac{\partial}{\partial t_1}\right)^k P_m\{t\}.
\label{difshur}
\ee

The object, defined by the r.h.s.\ of eq.~(\ref{infdet}), satisfies
the set of Hirota equations and thus is a Toda-lattice
$\tau$-function as a corollary of (\ref{difshift}). However, the
origin of the Hirota equations from this point of view is somewhat
more general than eq.~(\ref{difshift}).

\section{The origin of the bilinear equation}

The Hirota equations or bilinear identities have different
interpretations according to the different constructions which are used to
interpret the $\tau$-function:

1) From the point of view of integrable hierarchies these just
appear as a compact way to write the entire hierarchy in terms of
a single identity (\ref{Hirota}).

2) In the formalism of free fermions the simplest origin of
the bilinear identites is the formula for the quadratic Casimir operator of
the corresponding version of $GL(\infty)\otimes GL(\infty)$,
$\sum_{n=-\infty}^{\infty} \tilde\psi_n \otimes \psi_{-n}$,
which, together with the embedding of $GL(\infty)$ into the universal
enveloping of $\widehat{U(1)}_{k=1}$, straightforwardly implies
(\ref{Hirota}).

3) Eqs.~(\ref{hirdis1}), (\ref{hirdis2})
for vanishing multiplicities $p_a,p_b,p_c,
p,\bar p$ are immediate consequences of the determinant formula
(\ref{Miwadet}).
Eqs.~(\ref{hirdis1}), (\ref{hirdis2})
for arbitrary integer $p$'s can also be obtained in this way, if
$p_a>1$ is interpreted as the number (multiplicity) of coincident
"eigenvalues" $\lambda_a$. If $\lambda_b$ is smoothly moved along the spectral
surface and reaches  the position of $\lambda_a$, then $p_a$ changes abruptly:
$p_a\to p_a+p_b$. Within this context all vanishing $p_a$ can be considered
as associated with $\lambda_a=\infty$, and the change of $p_a$ by one
$p_a\to p_a+1$ is interpreted as "bringing" one extra "eigenvalue"
$\lambda_a$ from infinity.

4) The most general bilinear identities can be deduced by
the following procedure.  Introduce two
operators, $\hat M$ and $\hat{\bar M}$, acting on the infinite-dimensional
matrix
$H_{ij}$:
\be
\hat M : \  H_{ij} \longrightarrow H_{i+1,j}; \ \ \ \hat{\bar M} : \ H_{ij}
\longrightarrow H_{i,j+1}.
\label{shifop}
\ee
Select some $n\times n$ submatrix of $H_{ij}$ and consider
its determinant,
\be
\tau_{(n)}[\alpha,\bar\alpha | \beta,\bar\beta] =
\det_{{\alpha \leq i \leq \beta}\atop{\bar\alpha
\leq j \leq \bar\beta}} H_{ij}, \ \ n = \beta -\alpha =
\bar\beta -\bar\alpha.
\label{gentau}
\ee
Then $\hat M$ and $\hat{\bar M}$ act on $\tau_{(n)}$ by shifting the
$\alpha$'s and $\beta$'s:
\be
\hat M :\ \alpha \rightarrow \alpha +1,\ \beta \rightarrow \beta +1, \  \bar
\alpha \rightarrow \bar\alpha,\ \bar\beta \rightarrow \bar\beta; \nn \\
\hat {\bar M} :\ \alpha \rightarrow \alpha,\ \beta \rightarrow \beta, \  \bar
\alpha \rightarrow \bar\alpha +1,\ \bar\beta \rightarrow \bar\beta + 1,
\ee
and the following bilinear relation is true:\footnote{ It is
 a particular ($p=2$) case of the general identity for the minors of
any matrix,
\be
\sum_{i_p}H_{ri_p}\hat H_{i_1\ldots i_p|j_1\ldots j_p} = \frac{1}{p!}\sum_P
(-)^P
\hat H_{i_1\ldots i_{p-1}|j_{P(1)}\ldots j_{P(p-1)}} \delta_{rj_{P(p)}}, \nn
\ee
where the sum on the r.h.s.\ is over all permutations of the $p$
indices and $\hat H_{i_1\ldots i_p|j_1\ldots j_p}$ denotes the
determinant (minor) of the matrix, which is obtained from
$H_{ij}$ by removing the rows $i_1\ldots i_p$ and the columns
$j_1\ldots j_p$. Using the fact that $(H^{-1})_{ij} = \hat
H_{i|j}/\hat H$, this identity can be rewritten as
\be
\hat H \hat H_{i_1\ldots i_p|j_1\ldots j_p} =
\left(\frac{1}{p!}\right)^2\sum_{P,P'} (-)^P(-)^{P'}
\hat H_{i_{P'(1)}\ldots i_{P'(p-1)}|j_{P(1)}\ldots j_{P(p-1)}}
\delta_{i_{P'(p)}|j_{P(p)}}. \nn
\ee
Taking now $p=2$ and $i_1 = \alpha,\ i_2 = \beta,\ j_1 =
\bar\alpha,\ j_2 = \bar\beta$ one
obtains (\ref{HirJac}).  }
\be
\left( I\otimes \hat M \hat{\bar M} -
\hat M\otimes \hat{\bar M}\right)
\tau_{(n)}[\alpha,\bar\alpha | \beta,\bar\beta]
\tau_{(n)}[\alpha,\bar\alpha | \beta,\bar\beta]  =\\=
\tau_{(n+1)}[\alpha,\bar\alpha | \beta +1,\bar\beta +1]
\tau_{(n-1)}[\alpha +1 ,\bar\alpha +1 | \beta,\bar\beta].
\label{HirJac}
\ee

This formula is an identity, provided the action of $\hat M$
and $\hat{\bar M}$ is defined according to (\ref{shifop}). With
the same definition any two elements of the matrix $H_{ij}$ can
be related through $H_{ij} = \hat M^{i-i'} \hat{\bar M}^{j-j'}
H_{i'j'}$, or
\be\label{H}
H_{ij} = \hat M^i \hat{\bar M}^j H
\ee
(where $H = H_{00}$). Thus
\be
\tau_{(n)} [\alpha ,\bar\alpha |\beta, \bar\beta] =
\det_{{\alpha \le i < \beta}\atop{\bar\alpha \le j < \bar\beta}}
\left( \hat M^i \hat{\bar M}^j H\right).
\label{tauthrM}
\ee

One can now reformulate the statement and say that
(\ref{tauthrM}) is always a solution to (\ref{HirJac}), without
any references to the matrix $H_{ij}$ and the rules
(\ref{shifop}).  This is the general statement about the determinant
solution to the bilinear equations which we will need in this paper.

Let
us note, however, that this solution has the form
of the determinant of a {\it finite} matrix. Such objects are usually
interpreted as $\tau$-functions of {\it forced} hierarchies \cite{KMMOZ}.
Generic $\tau$-functions can be obtained from these through a limiting
procedure $\alpha$, $\bar \alpha\to \infty$:
\be
\tau_N=\lim_{n\to \infty}\tau_{(n)}[N-n,N-n|N,N].\nn
\ee

We will use
in the following both determinant representations since the infinite
matrix one is natural
from the viewpoint of the fermionic realization, while solution
(\ref{tauthrM}) is convenient to discuss algebraic properties.

It might be good at this point to explain, how the conventional
differential Hirota equation can be deduced from this general
construction. It arises from a particular choice of operators
$\hat M$ and $\hat{\bar M}$. Namely, consider a function
$F(X,\bar X)$ and define:
\be\label{29}
\hat M F(X,\bar X) = F(qX,\bar X), \ \ \ \hat{\bar M}F(X,\bar X) =
F(X,\bar q\bar X)\quad \mbox{(multiplicative action)}
\ee
or
\be\label{30}
\hat M f(x,\bar x) = f(x + \epsilon ,\bar x), \ \ \
\hat{\bar M}f(x,\bar x) = f(x,\bar x + \bar\epsilon )\quad
\mbox{(additive action).}
\ee
One can  also introduce
an algebraic version of the group action, specified by the operator
$\hat M$. It is defined in terms of the ``derivative''
operator\footnote{ Often a more symmetric definition $\hat
D = \hat\sigma(\hat M^{1/2} - \hat M^{-1/2})$ is used. It gives
expressions in a more symmetric form like the analogue of
(\ref{comforD}), $\left. \Delta(\hat D) = \hat D\otimes \hat
M^{1/2} + \hat M^{-1/2}\otimes \hat D\right|_{\rm diag}$, but
makes formulas somewhat more lengthy. Since this is not
important for our presentation, we stick here to the simpler definitions.  }
\be\label{defD}
\hat D = \hat\sigma (\hat M - I),
\ee
and similarly for $\hat{\bar D}$.
In the additive case $\hat\sigma$ simply acts  as multiplication by
a $c$-number,
$\hat\sigma = \epsilon^{-1}$. In the multiplicative
case $\hat\sigma$ is usually defined to act as
multiplication by $\frac{1}{(q-1)X}$,
(analogously, $\hat{\bar\sigma}$ is defined to act as multiplication by the
factor
$\frac{1}{(\bar q-1)\bar X}$) and has a non-trivial
commutation relation with $\hat M$:
\be
\hat\sigma \hat M = q\hat M\hat\sigma .
\ee

As a corollary of the above definitions, we have:
\be
\hat D^n = \left[\hat\sigma(\hat M - I)\right]^n =
\hat\sigma^n \prod_{k=0}^{n-1}\left( \hat M q^{-k} - I\right) =
\hat\sigma^n \hat M^n q^{-n(n-1)/2} \left( \hat M^{-1}; q\right)_n.
\ee
When acting on a product of two functions $F(X)G(X)$ (i.e.\ on the
``diagonal" of the tensor product) $\hat M$
is characterized by the following comultiplication property:
\be
\left. \Delta(\hat M) = \hat M \otimes \hat M \right|_{\rm diag}, \ \
{\rm i.e.} \nn \\
\hat M [F(X)G(X)] = F(qX)G(qX) = [\hat M F(X)][\hat M G(X)].
\ee
Since $\hat\sigma$ acts on the diagonal as multiplication,
\be
\left. \Delta(\hat\sigma)\right|_{\rm diag} = \hat\sigma,
\ee
we conclude that
\be
\left. \Delta(\hat D)\right|_{\rm diag} =
\left. \Delta\left(\hat\sigma(\hat M - I)\right) \right|_{\rm diag} =
\hat\sigma \left.\Delta(\hat M - I)\right|_{\rm diag} = \nn \\
= \hat\sigma \left.(\hat M\otimes \hat M - I\otimes
I)\right|_{\rm diag} =
\left. \left( \hat D \otimes I + I\otimes \hat D +
\hat \sigma^{-1} \hat D \otimes \hat D\right)\right|_{\rm diag} = \nn \\
= \left. \left(\hat D\otimes \hat M + I \otimes \hat
D\right)\right|_{\rm diag} .
\label{comforD}
\ee
Since $\hat\sigma^{-1} = (q-1)X$ vanishes at $q=1$, we obtain in this
``classical'' case the ordinary Leibnitz rule $\left.
\Delta(\hat D) = \hat D\otimes I + I\otimes \hat D\right|_{\rm diag}$.

The bilinear identity (\ref{HirJac}) can be, of course, rewritten in
terms of the operators $\hat D$ and $\hat{\bar D}$ instead of $\hat
M$ and $\hat{\bar M}$. Indeed, using the action of the
operators $\hat M$ and $\hat{\bar M}$ (\ref{29}), (\ref{30}), one can write:
\be
\tau_{(n)}(qX,\bar q\bar X)\tau_{(n)}(X,\bar X)-\tau_{(n)}(qX,\bar X)
\tau_{(n)}(X,\bar q\bar X)=
\tau_{(n-1)}(X,\bar X)\tau_{(n+1)}(qX,\bar q\bar X)
\ee
in the multiplicative variables, or
\be\label{simhir}
\tau_{(n)}(x+\epsilon,\bar x+\bar\epsilon)\tau_{(n)}(x,\bar x)-
\tau_{(n)}(x+\epsilon,\bar x)\tau_{(n)}(x,\bar x+\bar\epsilon)=\\=
\tau_{(n-1)}(x,\bar x)\tau_{(n+1)}(x+\epsilon,\bar x+\bar\epsilon)
\ee
in the additive variables. It can be equally rewritten in a form similar to
(\ref{HirJac}):
\be
\left( I\otimes \hat M \hat{\bar M} - \hat M\otimes \hat{\bar M}\right)
\tau_{(n)} \tau_{(n)} = I\otimes\hat{\bar M}
\hat M\tau_{(n-1)}\tau_{(n+1)},
\label{HirdifM}
\ee
or, using (\ref{defD}):
\be
\left( I\otimes \hat D \hat{\bar D} - \hat D\otimes \hat{\bar D}\right)
\tau_{(n)} \tau_{(n)} = I\otimes\hat{\bar \sigma}\hat{\bar M}
\hat\sigma\hat M\tau_{(n-1)}\tau_{(n+1)}.
\label{Hirdif}
\ee
More important, the same is true
for $\tau_{(n)}$ as given in (\ref{tauthrM}) (if it is considered as the
$\tau$-function of a
forced hierarchy -- see above). Indeed,
\be
\tau^{(q)}_N  = \det_{0\le i,j < N} \hat M^i\hat{\bar M}^j H =
\det (I + \hat\sigma^{-1}\hat D)^i (I +
\hat{\bar\sigma}^{-1}\hat{\bar D})^j H = \nn \\
= (q\bar q)^{(N-1)(N-2)/2}
\hat\sigma^{-N(N-1)/2}\hat{\bar\sigma}^{-N(N-1)/2} \det \hat D^i
\hat{\bar D}^j H.
\label{taudetdif}
\ee
One is actually free to choose an arbitrary normalization of the
$\tau$-function.
In particular, a natural choice, which cancels the $\sigma$-factors
in the r.h.s.\ of (\ref{Hirdif}) (compare with (\ref{hirdis1}),
(\ref{hirdis2})), is
\be\label{taudetdif2}
\hat\tau^{(q)}_N  = \det \hat D^i \hat{\bar D}^j H.
\ee
These normalized $\tau$-functions satisfy the equation
\be
\left( I\otimes \hat D \hat{\bar D} - \hat D\otimes \hat{\bar D}\right)
\hat\tau_{N} \hat\tau_{N} = \left(I\otimes\hat{\bar M}
\hat M\right)\hat\tau_{N-1}\hat\tau_{N+1}
\ee
which should be compared with (\ref{HirdifM}).
In the $q=1$ case, formula (\ref{taudetdif2})
is exactly the classical expression
of the theory of the Toda-lattice $\tau$-functions, included as
p.4) in the list of sect.~2. For $q\neq 1$ this result was first
derived in refs.~\cite{Sat} and interpreted there as the determinant
formula for solutions of a finite-difference Hirota equation.

\section{Introduction of higher time-variables}

There is still one ingredient missing in our discussion for
generic $q$ (we consider $|q|<1$ whenever it might lead to ambiguity).
The thing is that the operators $\hat M$ and $\hat D$
are defined as acting on functions of a single variable $X$.
Comparison with the $q=1$ case implies that $X$ plays the role
of the {\it first} time-variable $X = T_1$ (and $\bar X = \bar
T_1$ while $N$ can be considered as one of the zero-times).
According to this, (\ref{Hirdif}) is only the subset of
Hirota equations, associated with the first time-variables. A
way to introduce higher times is suggested by the relation (\ref{difshift})
for the $q=1$ case. A natural generalization is
\be
\hat D_k H = (\hat D_1)^k H, \ \ \ \hat{\bar D}_k H = (\hat{\bar D}_1)^k H,
\label{dkverd1}
\ee
where $\hat D_k$ is a finite-difference operator w.r.t.\ the
$k$-th time $T_k$,
\be
\hat D_k F(T_k) = \frac{F(q_kT_k) - F(T_k)}{(q_k-1)T_k}.
\ee
We reserve the possibility to take $q_k \neq q \equiv q_1$. For
example, the choice
\be
q_k = q^k
\ee
seems to be a reasonable alternative.

It was a corollary of (\ref{difshift}) that $H_{ij} =
\partial^i\bar\partial^j H = \partial_i\bar\partial_j H$
acquired the form (\ref{H-shur}), which is important for establishing the
relation to free fermions.
Let us now derive the analogue of (\ref{H-shur}) for $q\neq 1$.
The generic solution to eqs.(\ref{dkverd1}) is given by an integral
formula involving $q$-exponentials\footnote{
Let us remind the reader that the $q$-exponential
$e_q(x) = 1/E_q(-x)$ is characterized by the
following set of properties:

1. $D^+ \hat e_q(x) = \hat e_q(x)$;
$\hat e_q(x) \equiv  e_q\left((1-q)x\right)$ and
$\lim_{q \rightarrow 1} \hat e_q(x) = e^x$;

2. $e_q(x) = \sum_{k \geq 0} \frac{x^k}{(q,q)_k}$,
$E_q(x) = \sum_{k \geq 0} \frac{q^{k(k-1)/2} x^k}{(q,q)_k}$, \\
where $(a,q)_k \equiv \prod_{i=0}^{k-1} (1-aq^{i}) =
\frac{(a,q)_\infty}{(aq^k,q)_\infty}$;

3. $e_q(x) = \frac{1}{(x,q)_{\infty}}$, thus
$E_q(x) = (-x,q)_\infty$  and
$\theta_{00}(x) \equiv \sum_{k = -\infty}^{\infty}q^{k^2/2}x^k =
(q,q)_\infty E_q(q^{1/2}x)E_q(q^{1/2}x^{-1})$;

4. $E_q(x) E_q(y) = E_q(x+y)$ and
$e_q(y) e_q(x) = e_q (x+y)$, provided $xy = qyx$;

The first three properties explain the relevance of the $q$-special
functions as solutions to finite difference equations (i.e.\ to
various periodicity constraints), the last property indicates why these
same functions occur in the study of non-commutative
algebras and problems of quantum mechanics and quantum field theory.
}:
\be
H = \int dzd\bar z \left(\prod_{k=1}^\infty \hat e_{q_k}(T_k z^k)\right)
\mu(z,\bar z)
\left(\prod_{k=0}^\infty \hat e_{q_k}(\bar T_k \bar z^k)\right).
\label{Hint}
\ee
The measure $\mu(z,\bar z)$ is not specified by (\ref{dkverd1}).

Introduce now the $q$-Schur polynomials $P^{(q)}_n\{T\}$ through
\be
\prod_{k=0}^{\infty} \hat e_{q_k}(T_kz^k) = \sum_{n=0}^\infty
P^{(q)}_n\{T\}z^n.
\ee
It follows that $\hat D_k P^{(q)}_n\{T\} = P^{(q)}_{n-k}\{T\}$ and that
\be
H^{(q)}_{ij} = \hat D^i \hat{\bar D}^j H = \sum P^{(q)}_{l-i}
T_{lm} P^{(q)}_{m-j}, \nn \\
T_{lm} = \int_{dzd\bar z}z^l\mu(z,\bar z)\bar z^m.
\ee

The next question is that of the free-fermion representation of
$\tau^{(q)} = \det H^{(q)}_{ij}$. It could at first seem natural to look
for  a representation in terms of $q$-fermions, $q$-oscillators
etc. However it turns out that the most straightforward option is to
stay with the ordinary free fermions, the same as for $q=1$, and thus,
with the ordinary Segal-Wilson Grassmannian. (This phenomenon has already
been observed in ref.~\cite{MV}.)  The
origin of this possibility lies in the fact that $q$-Schur
polynomials and other relevant objects can be obtained from
their $q=1$ counterparts by a simple and universal change of
time-variables. Indeed,
\be
\prod_{k=1}^\infty \hat e_{q_k}(T_k z^k) = \prod_{k=1}^\infty e^{t_kz^k},
\ee
provided the $t$'s are expressed in terms of the $T$'s according to
\be
\sum_{k=1}^\infty t_kz^k = \sum_{n,k=1}^\infty
\frac{T_k^n(1-q_k)^n}{n(1-q_k^n)} z^{nk}.
\label{tverT}
\ee
Thus
\be
P^{(q)}_k\{T\} = P_k\{t\}
\ee
and
\be
H^{(q)}_{ij}\{T,\bar T\} = H_{ij}\{t,\bar t\}, \ \ \
\tau_N^{(q)}\{T,\bar T\} = \tau_N\{t,\bar t\},
\ee
where $H_{ij}$ satisfies eqs.~(\ref{difshift}), the entire set of
equivalences given in sect.~2 remaining valid.

Because of this, $\tau^{(q)}_N\{T,\bar T\}$ can be represented as
\be
\tau^{(q)}_N\{T,\bar T\} = \tau_N\{t,\bar t\} \ \stackrel{(\ref{fftau})}{=} \
\langle N | e^{H\{t\}} {\cal G} e^{\bar H\{\bar t\}} |N\rangle
\ee
with some ${\cal G} = \exp \left(\sum_{m,n}{\cal
A}_{mn}\tilde\psi_n\psi_n\right)$ and
\be\label{49}
H\{t\} = \sum_{n>0} t_nJ_{+n} \ \stackrel{(\ref{tverT})}{=} \
\sum_{n,k=0}^\infty \frac{T_k^n(1-q_k)^n}{n(1-q_k^n)} J_{+nk}, \nn \\
\bar H\{\bar t\} = \sum_{n>0} \bar t_n J_{-n} =
\sum_{n,k=0}^\infty \frac{\bar T_k^n(1-q_k)^n}{n(1-q_k^n)} J_{-nk}.
\ee

\section{Miwa transformation and discrete Hirota equations}

The bilinear identity for $q\neq 1$, when written in the form
(\ref{simhir}), is spectacularly similar to ``the discrete Hirota
equation'' (\ref{hirdis2}). That arises when the ordinary ($q=1$)
Hirota equation is represented in terms of  the Miwa coordinates.
This is perhaps a little less surprising once we know the relation
(\ref{tverT}), which, for the variable $T_1$, is nothing else than the Miwa
transformation (\ref{Miwatimes}). One difference is that
\be
t_k = \frac{1}{k}\frac{((1-q)T_1)^k}{1-q^k} =
\frac{1}{k}\sum_{l \ge 0} \left( (1-q)q^l T_1\right)^k
\label{semperMitr}
\ee
is essentially a simultaneous change of multiplicities
$p(q^{-l}\lambda_1)$ at all the points $q^{-l}\lambda_1$,
$l\geq 0$ on the spectral curve, rather than at a single point
$\lambda_1 = \left((q-1)T_1\right)^{-1}$. Another difference is that
eq.~(\ref{simhir}) is written in terms of the additive variables $x$, $\bar x$,
while $T_1$, $\bar T_1$ are associated with the multiplicative ones
$T_1=X$, $\bar T_1=\bar X$. We shall not go into a detailed discussion
of these differences here but will instead address another question.

What do the higher time-variables $T_k$ correspond to in this
context? Note first of all that there are generalizations of the
discrete Hirota equation (\ref{hirdis2}), corresponding to
simultaneous variation of  the multiplicities $p(\lambda)$ at several
arbitrary points $\lambda$ on the spectral curve, not necessarily
related by a ``semi-periodicity'' condition like (\ref{semperMitr}).
It is among these equations that we should
look for counterparts of the $T_k$ variations. In fact, the
relevant set of points $\lambda$, associated with the variable
$T_k$ is
\be
\left\{ \left.e^{2\pi i a/k}\lambda_k q_k^{-l/k}   \right| a=0,\ldots k-1;
\ \ l\geq 0\right\}, \ \ \
\lambda_k = \left( (1-q_k)T_k \right)^{-1/k}.
\ee
(Here we find a first reason to prefer the choice $q_k = q^k$
instead of $q_k = q$ since in the former case we have $q_k^{l/k}=q^l$.)
The action of $D/DT_k$ on $\tau^{(q)}_N$ can be now described as the
insertion of the
$k$-fermion non-local operator
\be
\Psi\left(\lambda_k e^{2\pi i a/k} \right) = \ :\prod_{a=0}^{k-1} \psi
(\lambda_k e^{2\pi i a/k}):
\ee
(note that only points with $l=0$ contribute):
\be
\hat M_{T_k} \tau^{(q)}_N \sim
\left< N+k\left| \Psi\left(\lambda_k e^{2\pi i a/k} \right)
e^{H\{t\}} {\cal G} e^{\bar H\{\bar t\}} \right|N\right> , \nn \\
\hat {\tilde M}_{\tilde T_k} \tau^{(q)}_N \sim
\left< N \left| e^{H\{t\}} {\cal G} e^{\bar H\{\bar t\}}
\bar \Psi\left(\bar\lambda_k e^{2\pi i a/k} \right)\right|N+k
\right>.
\ee

Thus, we conclude that the time-variables $T_k$ are associated with
certain types of Miwa transformations, distinguished by a particular
selection of points, where the multiplicities $p(\lambda)$ are
simultaneously changed.
One peculiarity is the ``semi-periodicity'' requirement,
forcing to introduce all the points $q^{-l}\lambda$, $0 \leq l <
\infty$ together with any $\lambda$. This is a usual thing in $q$-analysis
and $q$-free field theory (see, e.g.\ \cite{MV}), which can be easily
formalized in terms of Jackson integrals. Another peculiarity is the
association
of the multiples of $k$-th roots of unity with every $T_k$-variable. This
feature is, indeed, closely related to the notion of $k$-reduction of the KP
hierarchy, being, in fact, somewhat ``orthogonal'' to that of reduction
($k$-reduction
implies that the dependence on the variable $t_{kn}$ is completely {\it
eliminated}, while Miwa transformation, associated with $T_k$  does instead
{\it introduce} exactly this dependence and nothing else).

\section{Towards $q$-Matrix models}

The last promising direction for further research that will be mentioned
in this note concerns the search for ``$q$-matrix models''. We
restrict ourselves here to several preliminary remarks
concerning the relevant {\it eigenvalue} models and the associated
conformal theories \cite{UFN,Rev}, leaving the issue of matrices and
quantum groups for a more detailed presentation.

The partition functions of the eigenvalue models are usualy represented
in multiple integral form
\be\label{mm}
Z=\left[\prod_i \int_{dz_id\bar z_i}
\mu(z_i,\bar z_i)\right] \Delta(z)\Delta(\bar z)
 \prod_{i,k}e^{t_kz^k_i} \prod_{i,k}e^{\bar t_k\bar z^k_i}.
\ee
The measure $\mu(z,\bar z)$ in this expression depends on the concrete theory
and for the case of two-matrix model is equal to
$e^{-c(z-\bar z)^2}$.
This integral can be easily rewritten \cite{UFN,Rev,KMMOZ} as $\det H_{ij}$
with $H_{ij}=\int_{dzd\bar z}z^{i-1}\bar z^{j-1}\mu(z,\bar z)
\prod_{k}e^{t_kz^k} \prod_{k}e^{\bar t_k\bar z^k}$ which evidently satisfies
(\ref{difshift}) and, therefore,
$Z$ is a Toda-lattice $\tau$-function. The simplest model
arises when $c\to \infty$, i.e.\ $\mu(z,\bar z)=\delta(z,\bar z)$. It is
called the one-matrix model and corresponds to the reduction
of the Toda-lattice hierarchy to the Toda-chain one. The $\tau$-function of
this
hierarchy depends only on the difference of times $t_k-\bar t_k$ and
is generally described by a particular matrix $H_{ij}=H_{i+j}$.
The relevant matrix model integral is
\be\label{1mm}
Z^{(1mm)}=\left[\prod_i \int_{dz_i}\right] \Delta^2(z)
\prod_{i,k}e^{t_kz^k_i},
\ee
it gives rise to
\beq\label{Hij}
H_{ij}^{(1mm)}=H_{i+j}=\int_{dz} z^{i+j-2}\prod_{k}e^{T_kz^k}.
\eeq

There is an obvious way to deform these eigenvalue integrals. One
can just change all exponentials for $q$-exponentials. This is not
enough, however, to clarify the  connection with quantum groups.
The following remarks makes this relation a little clearer. The thing is
that the appearance of Van-der-Mondians in (\ref{mm})
is crucial for the representation of
such integrals in the form $\det H_{ij}$, characteristic of
$\tau$-functions associated to (forced) Toda-lattice
hierarchies. However, Van-der-Mondians are not very natural objects
in the theory of quantum groups. Their two natural analogues are either:

($i$) the quantum dimension which looks essentially like $\prod_{i>j}
(q^{z_i-z_j}-q^{z_j-z_i})$ \cite{KaLu} and is unsuitable if the deformed
integral still needs to be a Toda $\tau$-function, or

($ii$)
an object like
\beq\label{*}
\prod_{i>j}(qz_i-z_j)(z_i-z_j)
\eeq
which arises in the theory of $q$-free fields ($q$-affine algebras)
\cite{FJing}.
Remarkably, in the course of integration, it is possible to change variables
to replace (\ref{*})
by $\Delta^2(z)$ and and to obtain the appropriate determinant
formula. Indeed,
\be\label{58}
\left[\prod_i \int_{dz_i}\right] \prod_{i>j}(z_i-z_j)(qz_i-z_j)
\prod_{i,k}e^{t_kz^k_i}={\prod_{a=1}^{N}(1-q^a)\over (1-q)^N}\det H_{ij}=\\
={\prod_{a=1}^{N}(1-q^a)\over (1-q)^N}{\f N!}
\left[\prod_i \int_{dz_i}\right]
\prod_{i>j}(z_i-z_j)^2 \prod_{i,k}e^{t_kz^k_i}
\ee
with $H_{ij}$ defined in (\ref{Hij}).

The partition functions of the eigenvalue models can be usually represented
as correlators in two-dimensional conformal
field theory \cite{UFN,Rev}. This is especially useful for the study of
Ward identities. In particular, for the one-matrix model we have:
\be\label{cmm}
Z^{(1mm)}= \left[\prod_i \int_{dz_i}\right]
\left< e^{{1\over \sqrt{2}} \sum _{k>0}t_kJ_k}\prod_i e^{\sqrt{2}\phi(z_i)}
\right>_N,
\ee
where $\phi(z)$ is the free scalar field and $J(z)\equiv \partial \phi(z)$.

To deform this model in the way described above, one should perform the
substitution of times (\ref{49})
\be
\sum _{k>0}t_kJ_k \to \sum _{n,k>0}T_k^n{(1-q)^n\over n(1-q^n)}J_{nk}.
\ee
Note now that the square of the Van-der-Mondian in (\ref{1mm}) arises from the
operator product expansion of the $\widehat{sl(2)}_{k=1}$-currents
in (\ref{cmm}):
\beq
\prod_i e^{\sqrt{2}\phi(z_i)}=
\prod_i J^+(z_i) \sim \prod_{i>j}(z_i-z_j)^2.
\eeq
If these currents are replaced by those of the $\widehat{sl_q(2)}_{k=1}$
algebra \cite{FJing}, we get
\beq\label{62}
\prod_i J_q^+(z_i)\sim\prod_{i>j}(z_i-z_j)(qz_i-z_j).
\eeq
We thus obtain an argument in favor of the integrable system
(\ref{58}). The l.h.s.\ of (\ref{58}) is reproduced in this fashion only
for a particular choice of $q$-free fields. (It depends on the precise choice
of factors in front of the negative and positive harmonics of the field
$\phi(z)$; this can be done in different ways, without changing rules
like (\ref{62}).)

A very important ingredient in the theory of eigenvalue models
is the fact that one has explicit expressions for the Ward identities which
usually take the simple and recognizable form of Virasoro and
$W$-constraints.\footnote{ The fact that the full set of
Ward-identites (constraints) forms this kind of algebras is very
general, at least, as general as the existence of some free-field
representation of partition functions, in turn equivalent to
their identification as $\tau$-functions. It is remarkable that
in the simplest matrix models these algebras arise naturally
in some simple and well known representations.  }
In our $q$-matrix models, the Ward identities should presumably
obey some kind of $q$-Virasoro and $q$-$W$ constraints. We shall
not go into any details here, we shall merely note one important
identity, that arises if the choice
\be
q_k = q^k
\ee
is made. This choice respects the fact that the argument of the
$q$-exponential, while linear in $T_k$, involves the $k$-th power of $z$.
Because of this
\be
\frac{D}{Dz} \hat e_{q^k}(T_k z^k)  =
\frac{\hat e_{q^k}(T_kz^kq^k) - \hat e_{q^k}(T_kz^k)}{(q-1)} = \nn \\
= \frac{q^k-1}{q-1} T_k\frac{D}{DT_k}\hat e_{q^k}(T_kz^k).
\ee
This does not immediately lead to some simple formulas for the
constraints, because of the twisted nature of the $q$-Leibnitz
rule: in actual calculations, for models like (\ref{1mm}), every
$D/Dz_a$ will be acting on the product of $\hat e_{q^k}(T_k
z_a^k)$ with different $k$'s, while every $D/DT_k$ will be acting on the
product of $\hat e_{q^k}(T_k z_a^k)$ with different $a$'s. Integration
by parts is also somewhat more complicated than in the
$q=1$ case. To make this integration possible, one should certainly
use the adequate notion of integration, inverse to that
of finite-differentiation. It is essentially provided by the
Jackson integral,
\be
\int_q F(X) = (1-q)\sum_{l} F(q^l)q^l.
\label{Jack}
\ee
Usually the sum on the r.h.s.\ is defined over all
integers. Then, the integral of a total derivative
is given by the boundary terms,
$\int_q \hat DF(X) = F(\infty)-F(0)$. In order to get $q$-Virasoro
constraints, it would be useful to have the relation $\int_q \hat D
F(X) = 0$. For this purpose it will not be enough to simply use the integral
(\ref{Jack}), the reason being that it tends, in the classical limit, to an
integral over
the semi-axis $(0,\infty)$ and that there is no need for a reasonable function
$F(X)$ to vanish at zero.

There is however another integral which in the classical limit goes
to an integral over the entire real axis, namely,
\be
\oint_q F(X) = (1-q)
\sum_{l=-\infty}^{+\infty} \left\{F(q^l)+F(-q^l)\right\}q^l
\ee
and
\be
\oint_q\hat D F(X) = F(\infty)-F(-\infty).
\ee
A simple
example of a function, which decreases rapidly at both ends
of the integration domain (i.e.\ as $X\to \pm\infty$) is
$e_q(-X^2)=\prod_{k=0}^{\infty} (1 + q^kX^2)^{-1}$.

\section{Acknowledgements}

We are indebted to A. Gerasimov, S. Kharchev and A. Zabrodin for
numerous and stimulating discussions. A. Mironov and A. Morozov are also
grateful
to Centre de Recherches Math\'ematiques, Universit\'e de
 Montr\'eal for the kind hospitality. The work of A. Mironov is
partially supported by grant 93-02-14365 of the Russian
Foundation of Fundamental Research, that of L. Vinet through funds provided
by NSERC (Canada) and FCAR (Quebec).

\end{document}